\documentclass[12pt,preprint]{aastex}
\usepackage{graphicx}
\shorttitle{Eclipse of HMW 15}
\shortauthors{}

\begin{document}


\title{The Recurrent Eclipse \\of an Unusual Pre--Main-Sequence Star in IC 348}


\author{Stella Nordhagen\altaffilmark{1}}
\affil{Department of Physics and Astronomy, Middlebury College, Middlebury, VT 05753}
\email{snordhag@middlebury.edu}

\and

\author{William Herbst, Eric C. Williams}
\affil{Astronomy Department, Wesleyan University, Middletown, CT 06459}
\email{wherbst@wesleyan.edu, ewilliams@wesleyan.edu}

\and

\author{Evgeni Semkov}
\affil{Institute of Astronomy, Bulgarian Academy of Sciences, Sofia, Bulgaria}
\email{esemkov@astro.bas.bg}

\altaffiltext{1}{NSF-REU summer student at Wesleyan University}








\begin{abstract}

The recurrence of a previously documented eclipse of a solar-like pre--main-sequence star in the young cluster IC 348 has been observed. The recurrence interval is 4.7 $\pm\ 0.1$ yr and portions of 4 cycles have now been seen. The duration of each eclipse is at least 3.5 years, or  $\sim 75$\% of a cycle, verifying that this is not an eclipse by a stellar companion. The light curve is generally symmetric and approximately flat-bottomed. Brightness at maximum and minimum have been rather stable over the years but the light curve is not perfectly repetitive or smooth and small variations exist at all phases. We confirm that the star is redder when fainter. Models are discussed and it is proposed that this could be a system similar to KH 15D in NGC 2264. Specifically, it may be an eccentric binary in which a portion of the orbit of one member is currently occulted during some binary phases by a circumbinary disk. The star deserves sustained observational attention for what it may reveal about the circumstellar environment of low-mass stars of planet-forming age.

\end{abstract}

\keywords{binaries: eclipsing --- open clusters and associations: individual (IC 348)  --- stars: pre--main-sequence}


\section{Introduction}

Eclipses by non-stellar objects are of great interest for their potential to tell us not only about the stars involved but also about the extended objects -- presumably circumstellar or circumbinary disks. Such events signal their presence by their extended duration. A well known example is $\epsilon$ Aur, whose eclipse lasts  $\sim 2$ years  \citep{b85, l96}. Another is KH 15D \citep{kh}, an eccentric binary system embedded in an inclined circumbinary disk \citep{w06}. The longest known eclipse is for the star HMW 15\footnote{The position listed for this object by \citet{hmw00} is incorrect; the correct position is given by \citet{chw04}.}. This object, also known as TJ 108 \citep{tj}, H 187 \citep{h98} and LRLL 35 \citep{lrll}, is a member of  the young cluster IC 348 at a distance of about 300 pc \citep{c93, lrll, h07} and undergoes an eclipse that lasts $\sim 3.5$ years \citep{chw03} (hereafter Paper I). Like KH 15D it was discovered at Wesleyan University's Van Vleck Observatory (VVO) as part of a CCD photometric monitoring program of young clusters that has been going on for 15 years and covers about a thousand stars in several different clusters.

It was not clear from Paper I, which covered only five years of monitoring, whether this lengthy eclipse event would recur and, if so, on what time scale. Now that three additional years have passed, we have obtained enough data to answer these questions rather definitively. We have also obtained new color data that help to constrain models of the system. It is now clear to us that HMW15 is an object worthy of intensive study over the coming years and we hope that the new evidence for periodic recurrence of the eclipse will stimulate additional work on it by other investigators. 

\section{Observations and Reductions}

Observations of IC 348 at Wesleyan have been obtained since 1998; 
\citet{hmw00} and \citet{chw04} have reported results for the first five years, while the findings of the most recent study were documented in \citet{n06}. The interested reader is referred to those papers for a detailed description of the data acquisition and reduction methods used at VVO.  Six comparison 
stars (HMW 3, 7, 8, 10, 13, and 17) were employed in computing differential magnitudes and these  displayed standard deviations of less than 0.01 mag in each season. Instrumental differential
magnitudes (\emph{i}) were computed for each star on each night
relative to the average magnitude of comparison stars. These values
were transformed to standard magnitude (\emph{I}) using
the adopted offset of 11.23 mag, which provides an excellent match to the calibrated European system described below.
Differential magnitudes for HMW 15 were given in earlier papers in this series and an updated file is available upon request to the second author (W. H.).

Data were also obtained for the last two seasons with three European 
telescopes: the 2-m Ritchey-Chretien-Coude (RCC) and the 50/70 cm Schmidt (Scm)
telescopes of the National Astronomical Observatory Rozhen (Bulgaria) and
the 1.3-m Ritchey-Cretien telescope of the Skinakas Observatory\footnote{Skinakas Observatory is a collaborative project of the
University of Crete, the Foundation for Research and Technology - Hellas,
and the Max-Planck-Institut f\"{u}r Extraterrestrische Physik.} of the
Institute of Astronomy, University of Crete (Greece). These telescopes are equipped with the following makes of CCD, pixel sizes and scales, in respective order: VersArry (VA; 19 $\micron$; 0.25$\arcsec$/pix), Photometrics (Phot; 24 $\micron$; 0.5$\arcsec$/pix), and SBIG (ST8; 9 $\micron$; 1.1$\arcsec$/pix).
All frames were bias subtracted and flat fielded. CCD frames obtained with
the 50/70 cm Schmidt telescope were also corrected for dark counts. Twilight flat
fields in each filter were obtained on each night. All frames were taken
through a standard Johnson-Cousins set of filters. Aperture photometry was
performed using DAOPHOT routines. The typical exposure times are 60-120
sec for I, 120-180 sec for R and 180-300 sec for the V filter. There is generally excellent agreement between the data obtained at VVO and at the European observatories.

\section{Results}

\begin{figure}
\plotone{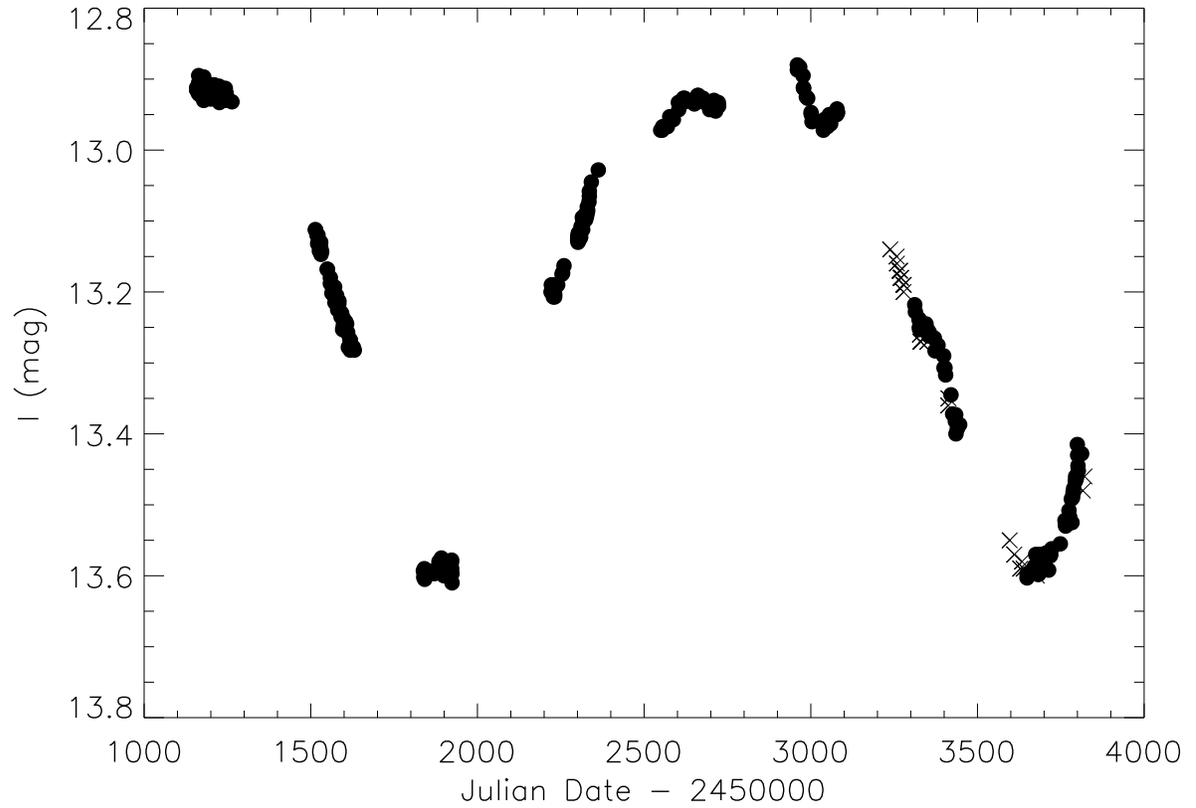}
\caption{Light curve for HMW 15 over the 8-year span of the Wesleyan observing program (solid circles), plus European data from Table 1 (crosses). Though the eclipse pattern seen in the 2nd-5th Wesleyan observing seasons appeared to break down in the 6th season, this proved to be a small deviation and the 7th and 8th seasons indicate that there is a repetition of the eclipsing behavior on a time scale of about 4.7 y.}
\label{lightcurve}
\end{figure}

The results of our previous study of HMW 15, based on data obtained over the first five years of the monitoring program and reported in Paper I, documented an apparent eclipse, unremarkable but for its extremely long duration. Relatively stable in the first season, the star faded by approximately 0.7 mag over the following season and remained stably at the fainter magnitude for the 2000-2001 season before recovering steadily to its original brightness throughout 2001-2002 and stabilizing during the 4th season. Although a few previous observations by \citet{tj} and \citet{h98} indicated the possibility of recurrence, the 2002-2003 season of data showed unexpected variation and no clear continuing pattern. However, the recently reduced data from the 2004-2006 seasons, plotted with the prior VVO observations in Figure \ref{lightcurve}, show a striking recurrence of nearly identical eclipsing behavior: the star fades steadily throughout the 2004-2005 season at nearly the same ingress slope observed in the 1999-2000 season, then, after (assumedly) remaining stable at the fainter magnitude for the summer of 2005, begins to brighten to its out-of-eclipse magnitude in the 2005-2006 season, again with a slope similar to that observed in the earlier egress. 

\begin{figure}
\plotone{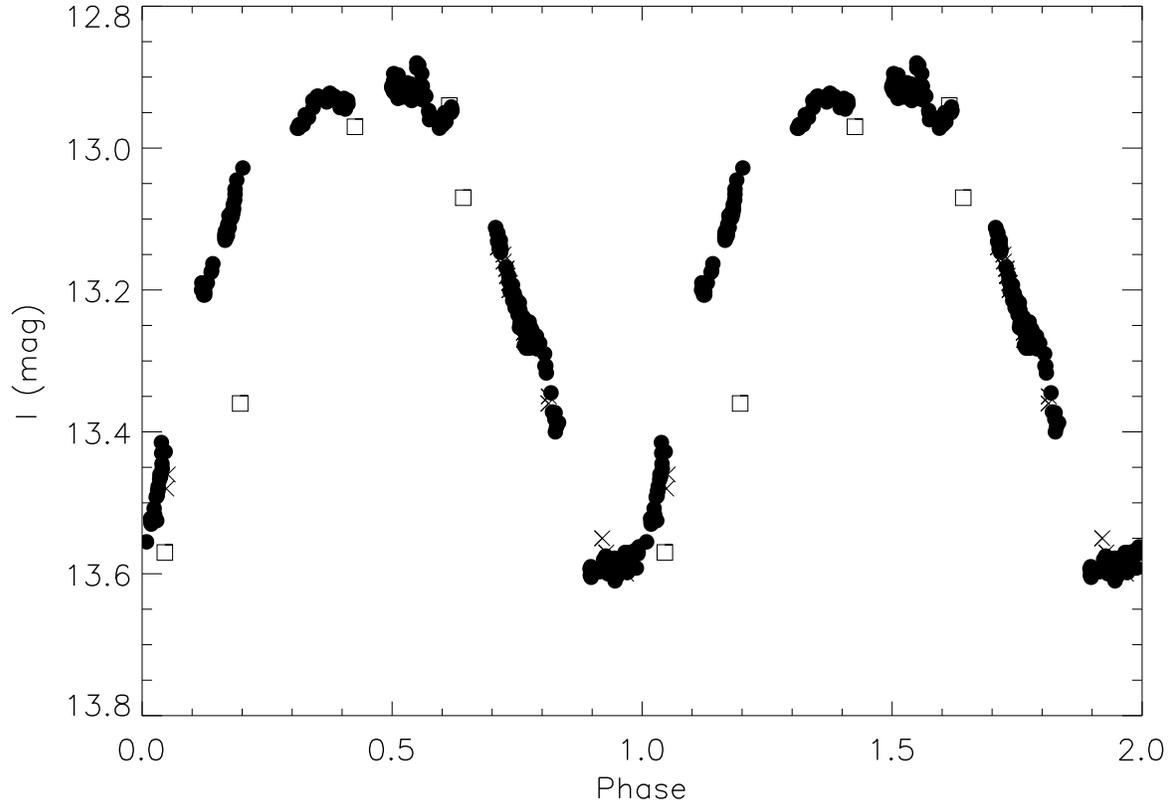}
\caption{Light curve for HMW 15 phased with a period of 4.7 y (1716.7 d). Solid circles are VVO data, crosses are European data from Table 1 and the open squares are data from 1992-1996 obtained by \citet{tj} and \citet{h98} as reported in Paper I. The one (open square) point which does not fit the observed pattern well is also the first data point obtained. Its deviation, if not a result of accidental error, indicates that there are stochastic or secular variations in the details of the light curve over time.}
\label{phase}
\end{figure}

Given the evident recurrence of the eclipse, the period of recurrence was estimated by visual  examination of phased light curves of different trial periods. Our ``best fit" phased light curve from this process is shown in Fig. \ref{phase} and is for a period of 4.7 y (1717 d). In addition to the recent data from VVO and Europe, we have included five earlier measurements by \citet{h98} and \citet{tj}, obtained in the early to mid-1990's. These data were reported in Paper I and they are shown on the plot as open squares. They probe one or two cycles earlier in the light curve and are, therefore, particularly important for period determination. 

As can be seen, our adopted period fits all of the data on this star well except for the very first measurement, which is from \citet{tj} and obtained on 24 October 1992. We can adjust the period somewhat and obtain a better fit for that one datum, but at the expense of a more scattered over-all fit. Given the fact that deviations from uniform behavior are clearly seen in the recent years, it is unclear how to interpret the deviation of that point. Perhaps there are secular changes in the shape of the light curve (similar to what is seen in KH 15D) or stochastic events that scatter the brightness at any phase. A good deal more data will be required to clarify this point. At present, we can simply say that a period of 4.7 $\pm$ 0.1 yr provides a good fit to essentially all the data, but that there is not perfect repeatability from cycle to cycle and there could be secular changes in the overall shape of the light curve. The brightness of the system near maximum and minimum light does seem to be rather consistent over the several cycles sampled so far. 

Armed with this period estimation and the knowledge that the eclipse phenomenon in this system has been a relatively stable event for at least the past 15 years, an attempt was made to determine whether or not the star's eclipsing behavior is a long term characteristic or a recent development. To this end, a search was made for the star on the Harvard Plate Collection (HPC), an impressively  extensive collection of astronomical photographic plates dating back to the late 19th century that has been used to unravel numerous similar stellar mysteries. Eighty-eight plates were examined, spanning the years 1889 to 1976, but unfortunately few plates approached the magnitude limits needed ($B \sim 17-18$, as the vast majority of the HPC plates are taken in $B$) to confirm the presence or absence of HMW15. Hence, we are unable to confirm whether the eclipses in this system were occurring prior to 1992.

It is clear from Figure \ref{lightcurve} that, in addition to the overall trend of the light, there are small scale deviations from uniformity, the most dramatic coming in the sixth season of observation when the overall fading of the star was replaced, for a period of a month or so, by a phase of rising brightness. Other deviations from uniform progression are visible at essentially all phases. There are not yet enough data to know for sure whether these are repeatable features of a strictly periodic light curve or stochastic events. We discuss possible sources of these features in the Discussion section. 

Finally, we show in Fig. \ref{color} the color behavior of the system during the last two seasons. Most of the data are in V-I so we concentrate on that. The circles represent the modern European data from this paper and the open squares are data obtained in 1993 and 1994 by \citet{tj} and \citet{h98}. The data reported here confirm previous indications that the star reddens as it fades, which is the normal behavior observed for TTS. This is now obvious and occurs in all colors. We also confirm that an interstellar reddening law (the solid line on Fig. \ref{color}) does not exactly fit the observed relationship, although it does not do too poorly. The data fall below the line indicating a ``flatter" than interstellar relationship, possibly indicative of reddening by  larger than typical interstellar grains. The dashed line is the prediction of a binary model discussed in the next section. 

To summarize this section, the results reported here confirm that the eclipse of HMW 15 is recurrent on a time scale of about 4.7 y. The light curve shape is not entirely smooth -- deviations from uniformity exist. The colors get redder as the star fades but not quite as rapidly as one would expect based on a standard interstellar extinction law. We turn now to an interpretation of these facts.

\section{Discussion}

HMW 15 is identified as a normal, PMS member of the young cluster IC 348 by \citet{lrll} based on its location on the sky and in a color-magnitude diagram. Its spectral type is listed as between G8-K4 in the optical and K3-K6 in the infrared by those authors, an unusually wide range of results although not uniquely so. We may infer that it is a relatively low mass star (0.5 to 1 M$_\odot$) at an age of 2-4 My and a distance of about 300 pc. It has no detectable infrared excess according to a recent {\it Spitzer} study by \citet{lm06} and is not known to be either a spectroscopic or visual binary. Its single distinguishing characteristic is its unusual photometric variability reported in Paper I and in this paper.

Our data demonstrate the importance of three time scales of variation for this star. One is the periodic time of 4.7 years, another is the eclipse duration time of about 3.5 years and a third is the ``wiggle" time of a few weeks during which coherent departures from smooth variation (sometimes involving reversals in the general trend of brightening or fading) occur. We discuss the plausible physical origins of each of these in turn.

The only plausible explanation for a period of several years, in our view, is that this is an orbital period. The dynamical time scale that would apply to either pulsation or rotation for a star like this is much shorter -- of order hours or days. If HMW 15 is a single star, then we surmise that it must be orbited by an inhomogeneous ring of matter with a period of 4.7 years, corresponding roughly to a semi-major axis of around 3 AU. Presumably this ring would be part of a more extensive structure, perhaps a circumstellar disk. In this model the 3.5 year eclipse time would indicate that for about one quarter of each cycle the material in this ring was either absent or of vanishingly small optical depth.  

Alternatively, and perhaps more plausibly, the period may be the orbital period of a binary star system. We imagine a system analogous to KH 15D during the 1960's-80's in which one member of a binary is obscured by a circumbinary disk during a portion of each cycle \citep{j04, j05}. As in the case of KH 15D this requires an eccentric orbit that is somewhat inclined to the plane of the circumbinary disk. The attractions of this model are the relatively flat bottom to the eclipse (expected if one star is totally obscured during that time) and the amplitude of about 0.75 mag, which is also expected for a total eclipse of stars with nearly the same luminosity. It should be easy to test this model over the next few years since radial velocity variations of the star or stars should easily be detectable. In this model, the 3.5 year extent of the eclipse is set by the fraction of the orbit covered by the circumbinary disk. 

Regardless of whether the star is single or a binary, it remains something of a puzzle to explain the shorter time scale variations that superimpose themselves on the overall light curve as small deviations or ``wiggles". One possibility is that they represent actual variations in the brightness of the star or stars due, perhaps, to starspots, which are ubiquitous on pre-main sequence stars. However, the normal time scale for such variability is much shorter than observed -- typically days, due to the rotation period of the star. Another possibility is that these represent irregularities in the optical depth of the screen -- lumps and clumps in the distribution of matter in the circumstellar or circumbinary disks in the models discussed above. We need to have a longer period of monitoring to see how frequently these events occur and to be able to describe their properties statistically before more can be said about them. Color information during the events would also be interesting to obtain.

\begin{figure}
\plotone{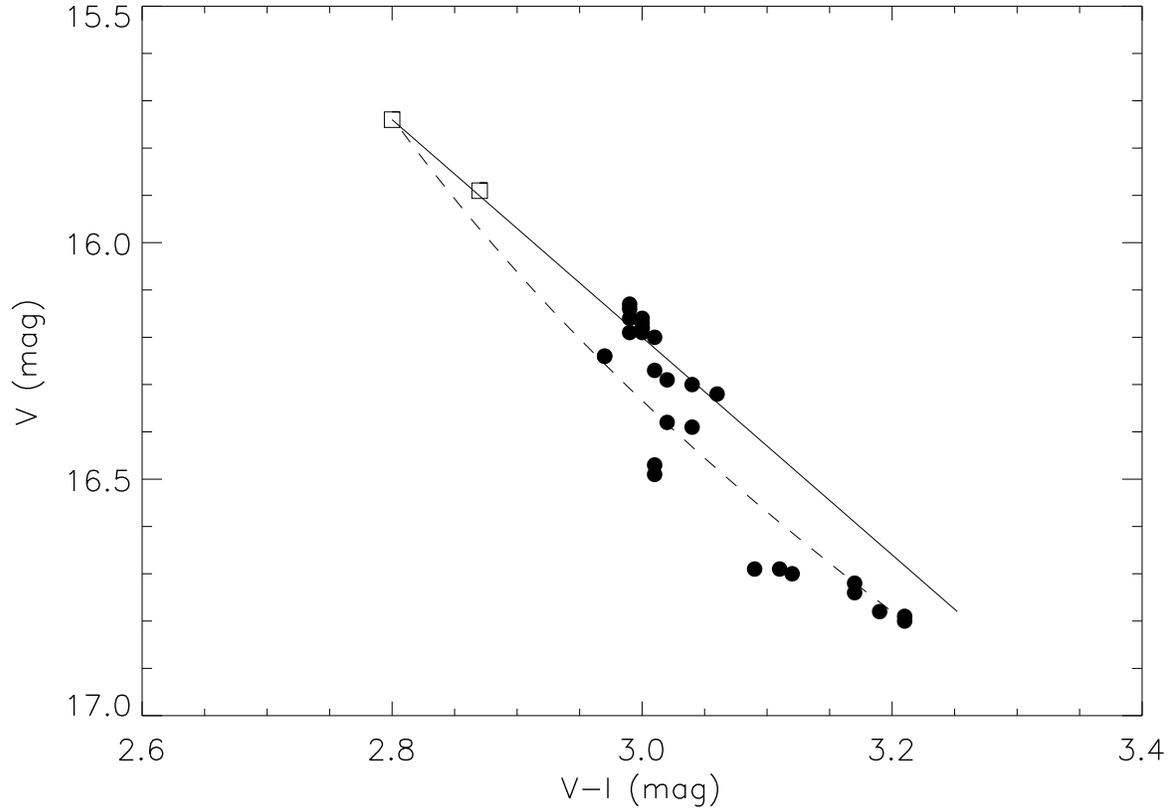}
\caption{Color behavior of HMW 15 with brightness from the European data (solid circles) and measurements (open squares) by \citet{tj} and \citet{h98}. Clearly the star reddens as it fades, in agreement with earlier indications summarized in Paper I. The solid line shows the standard interstellar reddening slope from \citet{t86}. The dashed line is for a binary model described in the text in which the bluer star in a G8 + K6 binary is progressively occulted by an opaque screen.}
\label{color}
\end{figure}

The color data do not allow us to distinguish between the single star or binary model for this object. While an interstellar reddening law does not fit the color data very well (solid line on Fig. \ref{color}) it is also not a terrible fit and it is always possible, perhaps even expected, that grain growth would have occurred in a circucmstellar disk of this age, so that a flatter than normal extinction law might be quite usual. On the other hand, a binary model can fit the data equally well, if not better, with very little effort. As an example, we show (dashed line on Fig. \ref{color}) the predicted color behavior for a G8 + K6 system in which the G8 star is entirely visible at maximum and entirely obscured at minimum light. In computing this model we have assumed the extinction is gray. If, instead, we allowed the extinction to follow an interstellar reddening line, then the system light would follow the solid line during its brighter part and the dashed line during its fainter part. Obviously this would be an even better fit to the data. 

The simplest binary model, with gray extinction and total obscuration of the bluer star, is unlikely to be correct for another reason. The required color difference between the components is quite large for their magnitude difference. If, as one would expect, the components of this putative binary were coeval, then they would be expected to lie along an isochrone in the V, V-I plane. The simple binary model described here does not come close to either a theoretical or observed isochrone. The redder component is too bright by about 1.5 mag in V. This probably means that we need to consider non-gray extinction models. If the binary model is correct and we can obtain the spectral types of both stars, then it should be easily possible to determine the extinction law and the intrinsic properties of the stars. In this case, the system should prove to be a useful test of PMS models since we will have coeval stars of different masses. At present, more refined models are not warranted given the meager available data.

It is interesting that HMW 15 escaped detection as a variable star for so long, in spite of the fact that its amplitude is more than one magnitude in V. Part of the reason is that it varies so slowly. Most variability studies, especially those aimed at young stellar objects, are tuned for detections on time scales of hours to months. While in some seasons this star is detectable as variable over a few weeks, it is not highly variable on that time scale and does not easily emerge from the data until multiple seasons are put together. Perhaps there are more such long time scale, periodic variables in young clusters awaiting detection. If so, we should continue to find them at VVO as the baselines for our variability studies continue to grow.

To conclude, our data do not rule out a single star model for this system, but the binary model is marginally more attractive since it does a better job of explaining the color behavior and perhaps the wide range in reported spectral types \citep{lrll}. The issue should soon be decided by spectral studies. If the star is a binary with an orbital period of 4.7 y and G or K components, then the radial velocity amplitudes of the stars will be easily detectable (many km/s). Regardless of this, it is now clear that at least one star in the system is periodically eclipsed by extended dust in its vicinity, presumably either a circumstellar or circumbinary disk. As in the case of KH 15D, we have a rare opportunity to learn about possibly planet-forming material by using the starlight as a probe. The relative motions of the star and extended matter provide a time dimension that can be exploited by a determined spectroscopic monitoring program supplementing continued photometry. We recommend this as a promising approach toward improving our knowledge of young stars and disks.

\acknowledgments We thank J. Winn of M.I.T. for helpful discussions about the star and he and A. Doane at Harvard for their assistance in searching the HPC . Additionally, we thank the numerous Wesleyan students, particularly Roger Cohen, who carried out eight years of photometric observations. S.N.'s work was supported by the National Science
Foundation under Grant No. AST-0353997 to Wesleyan University,
supporting the Keck Northeast Astronomy Consortium. This material is based upon work supported by the National Aeronautics and Space Administration under Grant NNG05GO47G issued through the Origins of Solar Systems Program to W. H. The authors thank the Director of Skinakas Observatory
Prof. I. Papamastorakis and Prof. I. Papadakis for some of the telescope time.

\begin{deluxetable}{cccccccc}
\tablewidth{0pt} 
\tablecaption{Photometry of HMW 15 at Rozhen and Skinakas Observatories}
\label{table1}  
\tablehead{\colhead{ JD (2453 ...)} & \colhead{I} & \colhead{R} & \colhead{V} & \colhead{B} & \colhead{U} & \colhead{telescope} & \colhead{CCD}}
\startdata
238.527 & 13.14 & 14.67 & 16.13 & 18.13 & 19.6 & 1.3m & Phot \\
257.514 & 13.15 & 14.67 & 16.14 &   & - & 1.3m & Phot \\
258.488 & 13.16 & 14.69 & 16.16 & - & - & 1.3m & Phot \\
266.589 & 13.17 & 14.69 & 16.16 &   & - & 1.3m & Phot \\
268.617 & 13.17 & 14.70 & 16.17 &   & - & 1.3m & Phot \\
269.534 & 13.18 & 14.70 & 16.17 &   & - & 1.3m & Phot \\
271.614 & 13.18 & 14.71 & 16.18 &   & - & 1.3m & Phot \\
277.440 & 13.19 & 14.74 & 16.20 &   & - & 1.3m & Phot \\
278.414 & 13.20 & 14.74 & 16.19 &   & - & 1.3m & Phot \\
279.426 & 13.19 & 14.73 & 16.19 & - & - & 1.3m & Phot \\
327.518 & 13.26 & 14.85 & 16.27 & - & - & Scm & ST8 \\
328.548 & 13.27 & 14.85 & 16.24 & - & - & Scm & ST8 \\
330.468 & 13.27 & 14.86 & 16.24 & - & - & Scm & ST8 \\
348.417 & 13.26 & 14.86 & 16.30 & - & - & Scm & ST8 \\
349.392 & 13.26 & 14.87 & 16.32 & - & - & Scm & ST8 \\
350.309 & 13.27 & 14.88 & 16.29 & - & - & Scm & ST8 \\
412.285 & 13.36 & 14.97 & 16.38 & - & - & Scm & ST8 \\
413.238 & 13.35 & 14.97 & 16.39 & - & - & Scm & ST8 \\
596.580 & 13.55 & 15.19 & 16.72 & - & - & 1.3m & Phot \\
610.486 & 13.57 & 15.20 & 16.74 & 18.83 & - & 1.3m & Phot \\
628.458 & 13.59 & 15.23 & 16.78 & 18.90 & - & 1.3m & Phot \\
633.513 & 13.58 & 15.23 & 16.79 & - & - & 1.3m & Phot \\
638.502 & 13.59 & 15.24 & 16.80 & - & - & 1.3m & Phot \\
678.276 & 13.60 & - & 16.69 & - & - & 2m & VA \\
700.295 & 13.58 & 15.27 & 16.69 & - & - & Scm & ST8 \\
701.316 & 13.58 & 15.28 & 16.70 & - & - & Scm & ST8 \\
816.294 & 13.48 & 15.06 & 16.49 & - & - & 2m & VA \\
821.276 & 13.46 & - & 16.47 & - & - & 2m & VA \\
\enddata
\end{deluxetable}



\clearpage

\clearpage

\clearpage

\clearpage




\end{document}